\documentclass{PoS}

\usepackage[square,sort&compress,numbers]{natbib}
\usepackage{subfig}
\usepackage{amsmath}
\usepackage{lineno}

\title{Odd and Even Partial Waves of $\boldsymbol{\eta \pi^-}$ and $\boldsymbol{\eta'\pi^-}$ in $\boldsymbol{191\,\textrm{GeV}/c}$ $\boldsymbol{\pi^-p}$}

\ShortTitle{Odd and Even Partial Waves in $\eta'\pi$ and $\eta\pi$}

\author{\speaker{Tobias Schl\"uter}\thanks{The author acknowledges
    financial support by the Maier-Leibniz-Laboratorium der LMU und TU
    M\"unchen, by the German Bundesministerium f\"ur Bildung und
    Forschung (BMBF), and by the DFG cluster of excellence ``Origin
    and Structure of the Universe.''}\\
  Ludwig-Maximilians-Universit\"at M\"unchen, Germany\\
  E-mail: \email{tobias.schlueter@physik.uni-muenchen.de}\\
  on behalf of the COMPASS collaboration}


\abstract{In the year 2008 COMPASS recorded diffractive events of the
  signature $\pi^-(191\,\textrm{GeV})p \to X_{\textrm{fast}} p$.  We
  present results of the analysis of the subsystems
  $X=\eta^{(\prime)}\pi^-$.  Besides the known resonances $a_2(1320)$,
  $a_4(2040)$, we study the properties of the spin-exotic $P_+$ wave,
  and all other natural-exchange partial waves up to spin $J=6$.  We find
  a striking difference between the two final states: whereas the even
  partial waves 2, 4, 6 in the two systems are related by phase-space
  factors, the odd partial waves are relatively suppressed in the
  $\eta\pi^-$ system.  The relative phases between the even waves
  appear identical whereas the phase between the $D$ and $P$ waves
  behave quite differently, suggesting different resonant and
  non-resonant contributions in the two odd-angular-momentum systems.
  Branching ratios and parameters of the well-known resonances $a_2$
  and $a_4$ are measured.  We find
  \[
    \label{eq:1}
    m(a_2) = 1315\pm 12\,\textrm{MeV}, \quad \Gamma(a_2) = 119\pm 14\,\textrm{MeV},
  \]
  and
  \[
    \label{eq:2}
    m(a_4) = 1900^{+80}_{-20}\,\textrm{MeV}, \quad
    \Gamma(a_4) = 300^{+80}_{-100}\,\textrm{MeV}.
  \]
  (consistent with COMPASS's $3\pi$ analyses.)  For the relative branchings we measure
  \[
    \frac{BR(a_2\to \eta'\pi)}{BR(a_2\to \eta\pi)} = (5\pm 2)\%,\quad \frac{BR(a_4\to \eta'\pi)}{BR(a_4\to \eta\pi)} = (23\pm 7)\%.
  \]
}

\FullConference{XV International Conference on Hadron Spectroscopy-Hadron 2013\\
		4-8 November 2013\\
		Nara, Japan }

\begin{document}

\section{Introduction}

The systems $\eta\pi$ and $\eta'\pi$ are attractive laboratories for
strong-interaction physics because of their simplicity and clear
experimental signature.  Besides the well-known resonances $a_2(1320)$
and $a_4(2040)$, resonance-like behavior was observed in the $P$-wave,
whose neutral isospin member carries the exotic quantum numbers
$J^{PC}=1^{-+}$ (see e.g.\ Ref.~\cite{Klempt:2007cp}).  In this contribution, we
discuss an analysis of the $\eta\pi^-$ and $\eta'\pi^-$ systems,
diffractively produced off a proton target during the 2008 run of the
COMPASS experiment.  Previous work on this analysis was discussed in
Refs.~\citep{Schluter:2012re,Schluter:2012}.  A journal publication is
in progress.

\begin{figure}[bh]
  \centering
  \subfloat[$\eta\pi^-$ Acceptance]{\includegraphics[width=.46\textwidth]{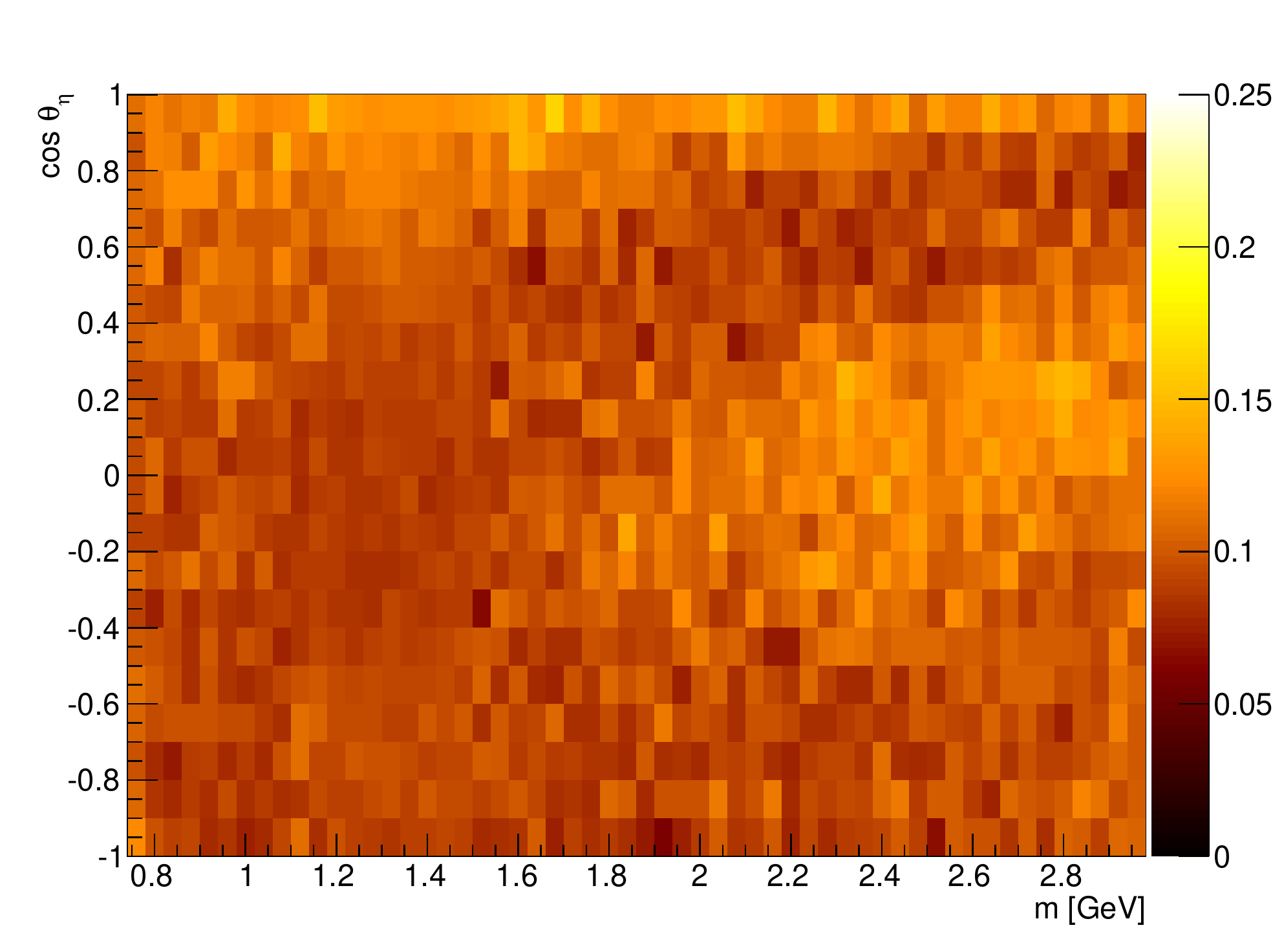}}\quad
  \subfloat[$\eta'\pi^-$ Acceptance]{\includegraphics[width=.46\textwidth]{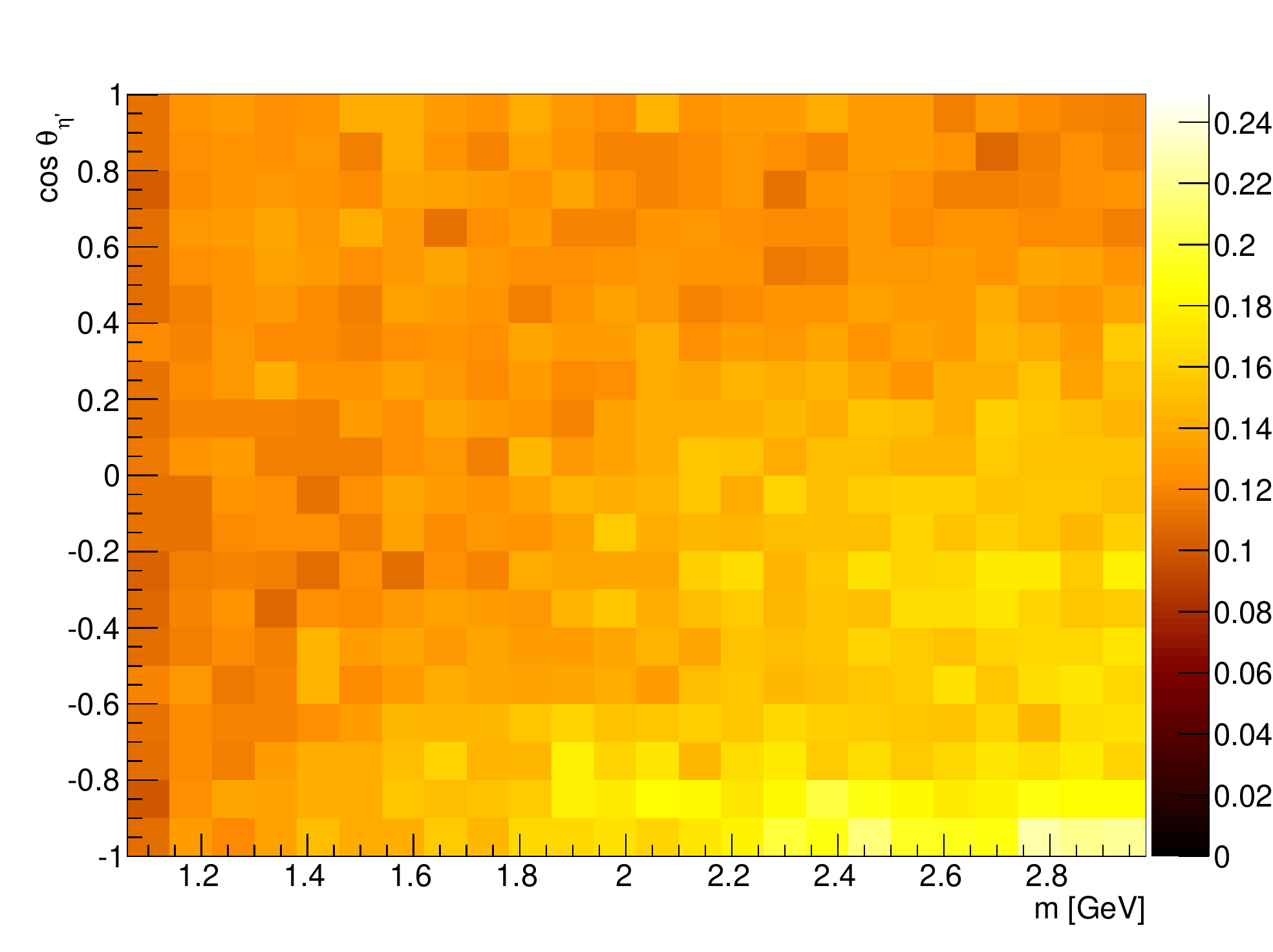}}
  \caption{Acceptance evaluated from Monte Carlo.  An azimuthal
    distribution $\propto \sin^2\phi$ (i.e.\ natural exchange, $M=1$) and
    the experimental $t'$ distribution were used for these pictures.}
  \label{fig:acceptance}
\end{figure}

The COMPASS experiment is a fixed-target experiment installed at the
CERN SPS.  Its two-stage spectrometer allows for high-resolution
particle detection and reconstruction over a wide range in angles and
momenta, both for charged and neutral particles~\cite{Abbon:2007pq}.
The data recorded for the analysis under discussion was produced by
having a $191\,\textrm{GeV}$ $\pi^-$ beam impinge on a LH2 target.
The target was surrounded by a recoil proton detector which together
with a veto detector surrounding the spectrometer entry formed a
trigger ensuring a clean sample of diffractive excitation reactions
with momentum transfer $|t| \gtrsim
0.08\,\textrm{GeV}^2$~\cite{Schluter:2011}.  Samples of approximately
$35\times 10^3$ exclusive $\pi^-\eta'$ and $110\times 10^3$ exclusive
$\pi^-\eta$ events with invariant masses from threshold up to several
$\textrm{GeV}$ is obtained in the reaction $\pi^-p\to
\pi^-\pi^-\pi^+\gamma\gamma p$, where the two photons result from the
decay of an intermediate $\eta$ ($\pi^0$) in the $\eta'$ ($\eta$)
decay.  Acceptances for the two reactions as function of mass and
polar angle (Gottfried-Jackson system) are shown in
Fig.~\ref{fig:acceptance}.

In the flavor basis, $\eta$-$\eta'$ mixing is described by an angle
$\phi \approx 39^{\circ}$~\cite{Beringer:2012zz}.  One expects in
particular for branching ratios of the $a_2$ and $a_4$ resonance
decays to pseudoscalars
\[
\textrm{BR}(a_J\to\pi\eta')/
\textrm{BR}(a_J\to\pi\eta)
 =
 F(J, q'(m), q(m)) \tan^2\phi,
\]
where $J$ is angular momentum, and $q^{(\prime)}(m)$ are the breakup
momenta at invariant mass $m$.  For $q^{(\prime)}\to 0$ the behavior
of each cross-section has to follow $(q^{(\prime)})^{2J+1}$ from
analyticity of the partial-wave series.  Therefore, the simplest form
of dynamical term, which we use in the following, is
\begin{equation}
  \label{eq:scale}
  F(J,q',q) = (q'/q)^{2J+1}.
\end{equation}

\section{Partial-wave Analysis Procedure}

\begin{figure}
  \subfloat[$p(\pi^+)$]{\includegraphics[width=.3\textwidth]{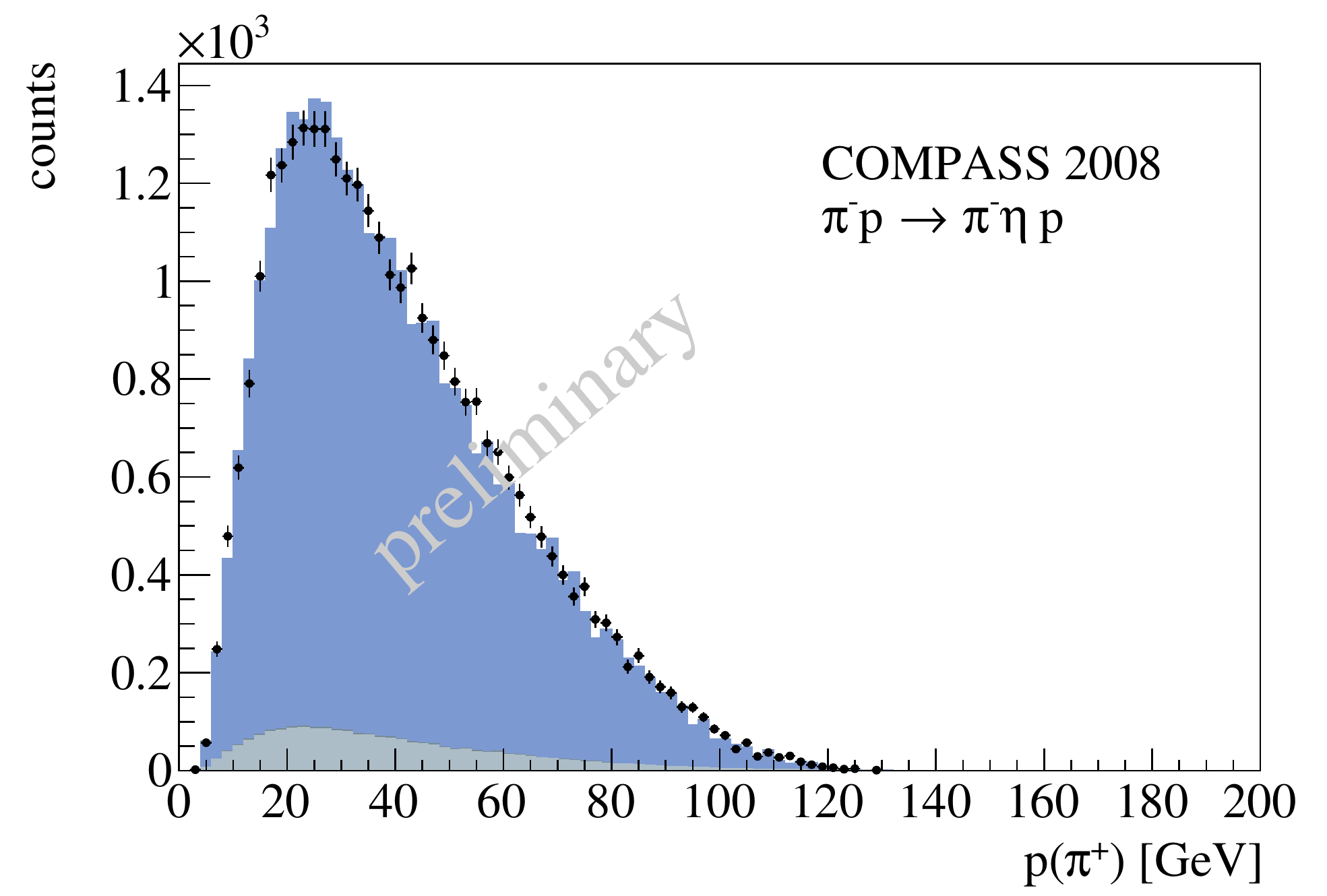}}\quad
  \subfloat[$p(\pi^-)$ (2 entries p.ev.)]{\includegraphics[width=.3\textwidth]{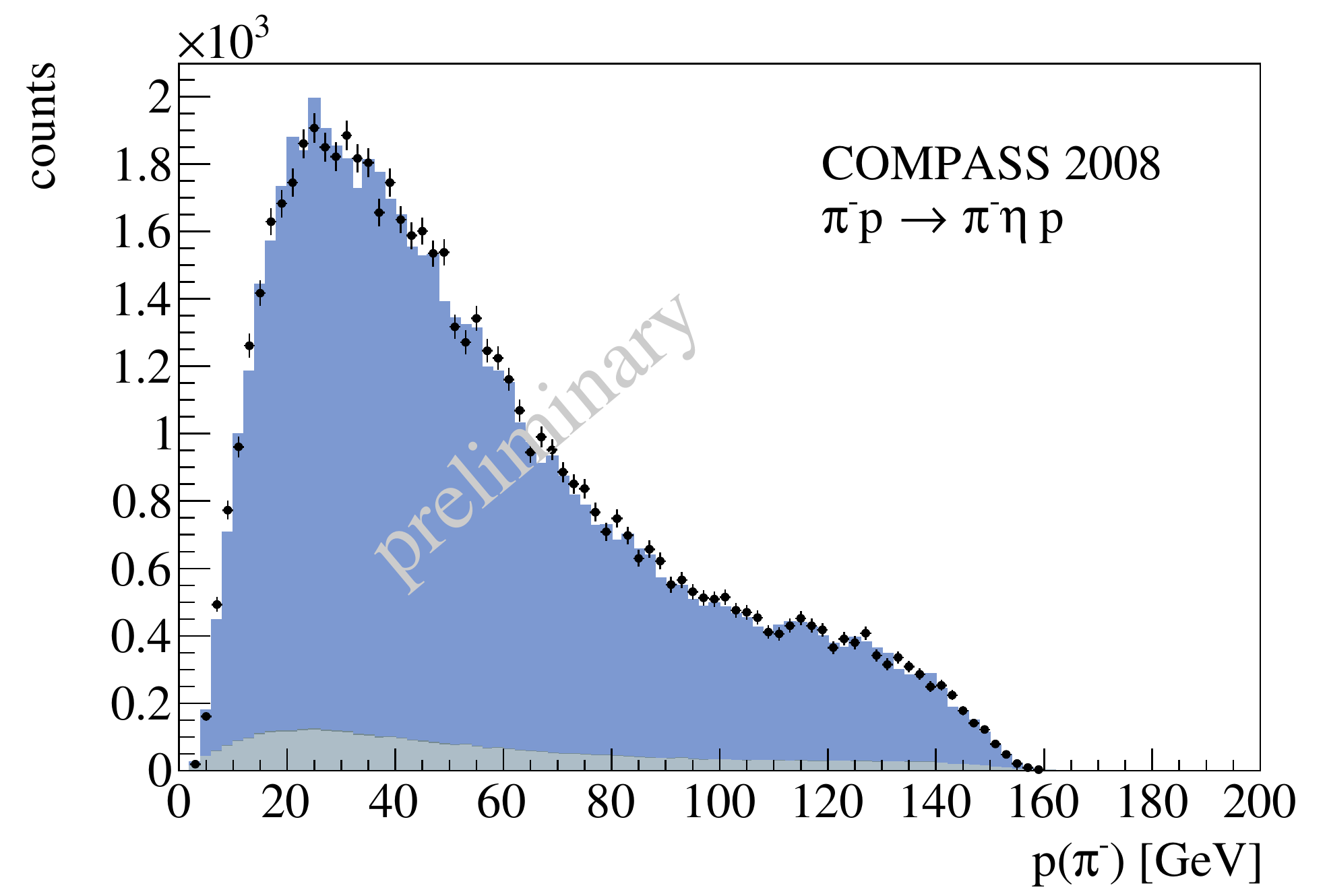}}\quad
  \subfloat[$p(\pi^0)$]{\includegraphics[width=.3\textwidth]{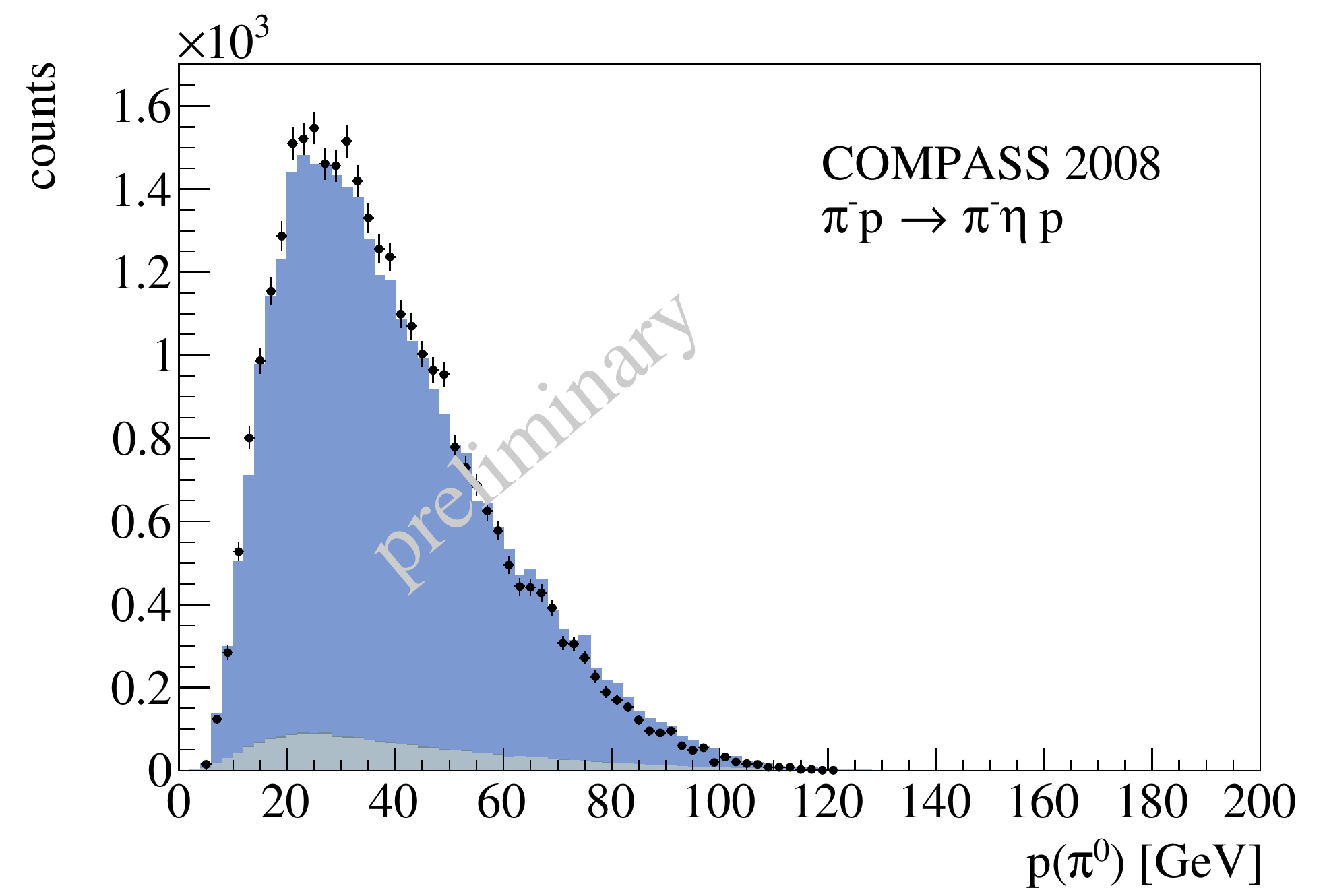}}\\
  \subfloat[$\phi_{\textrm{\tiny GJ}}$]{\includegraphics[width=.3\textwidth]{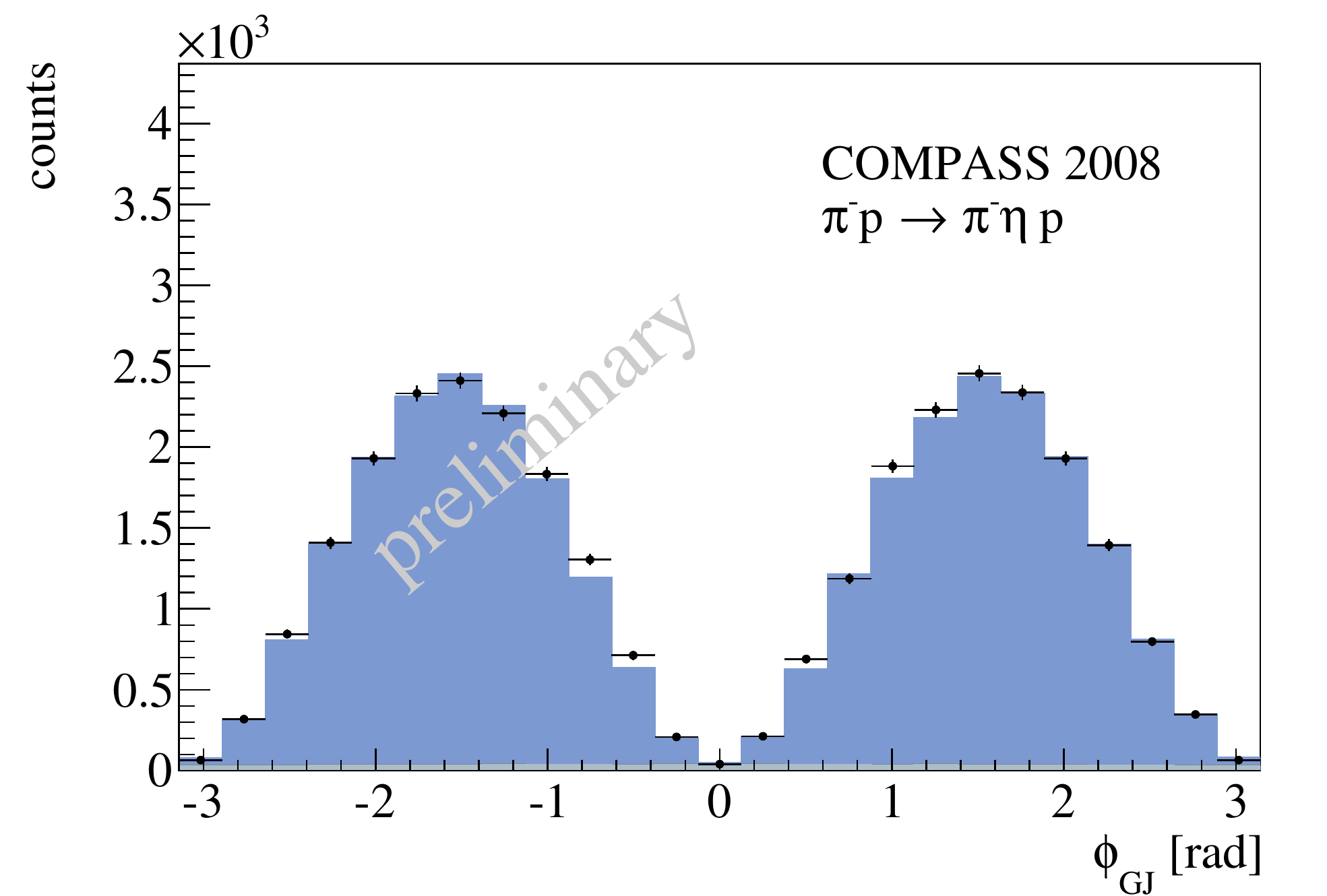}}\quad
  \subfloat[$\cos \theta_{\textrm{\tiny GJ}}$]{\includegraphics[width=.3\textwidth]{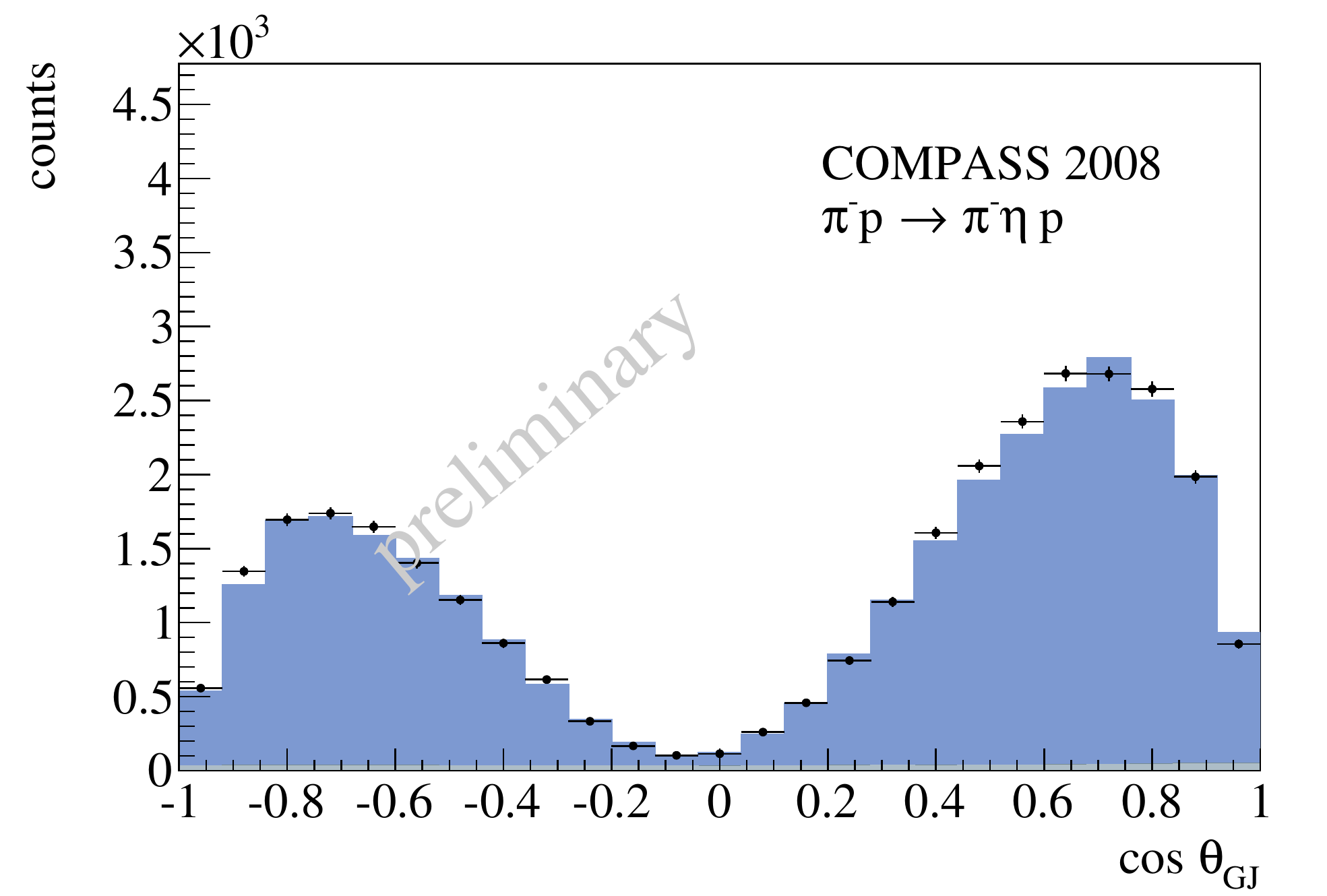}}\quad
  \subfloat[$m(\pi^-\pi^+\pi^0)$ (two entries p.ev.)]{\includegraphics[width=.3\textwidth]{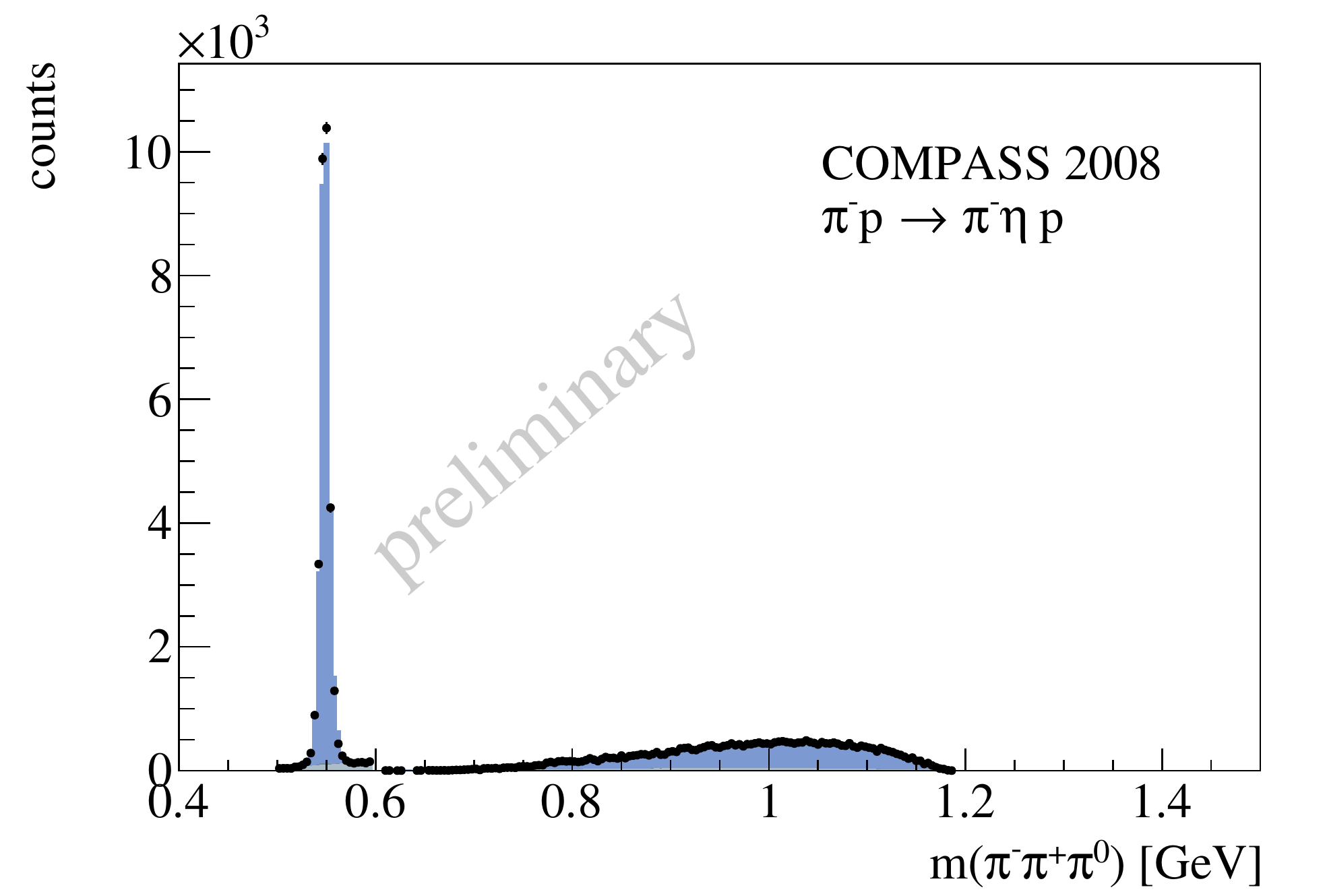}}

  \caption{Comparison of fit result to the data.
    First row: lab momenta of pions, second row: two-body angular
    variables and invariant three-body mass with $\eta(548)$ peak.
    Black dots: data.  The fit result is decomposed into the
    incoherent contributions as follows: Light blue: natural-exchange
    waves.  Gray: non-$\eta$ background.  Dark blue:
    unnatural-exchange waves (negligible).}
  \label{fig:distrib-predict}
\end{figure}

The data were subjected to partial-wave analysis.  Here, an
acceptance-corrected partial-wave model was fit to the data in
$40\,\textrm{MeV}$ wide bins of $m(\eta^{(\prime)}\pi^-)$ from
threshold up to $3\,\textrm{GeV}$, separately for the two final
states.  The formalism used was an extended $\log$-likelihood fit
where partial waves were parametrized in the reflectivity basis.
Natural-exchange partial waves with $M=1$ for angular momenta up to
$J=6$ were included.  For $J=2$ in $\eta\pi$, an $M=2$ partial wave
was also included in the analysis.  It was found to contribute $3\%$
of the $a_2$ intensity.  The complete four-body information was used
to distinguish the three-body $\eta^{(\prime)}$ peak from background
reactions which were modeled by a partial wave isotropic in four-body
phase space.  The partial wave model consisted of three incoherent
contributions: natural-exchange waves, unnatural-exchange waves and
the flat four-body background.  Consistent with the expectation of a
dominant Pomeron contribution, unnatural parity exchange is found to
be suppressed.  In Fig.~\ref{fig:distrib-predict} we illustrate the
procedure by overlaying the data and the fit results for the
$\eta\pi^-$ data in the vicinity of the $a_2(1320)$ resonance.  For
details see Ref.~\cite{Schluter:2012}.  Physical hypotheses are then
tested by fitting mass-dependent (resonance) models to the
partial-wave intensities and phases extracted in the mass-binned fit.
The Breit-Wigner fit results for the known resonances are given in the
abstract.

\section{Partial-wave Results}

\begin{figure}
  \centering
  \subfloat[$P$-wave, $J=1$]{\includegraphics[width=.32\textwidth]{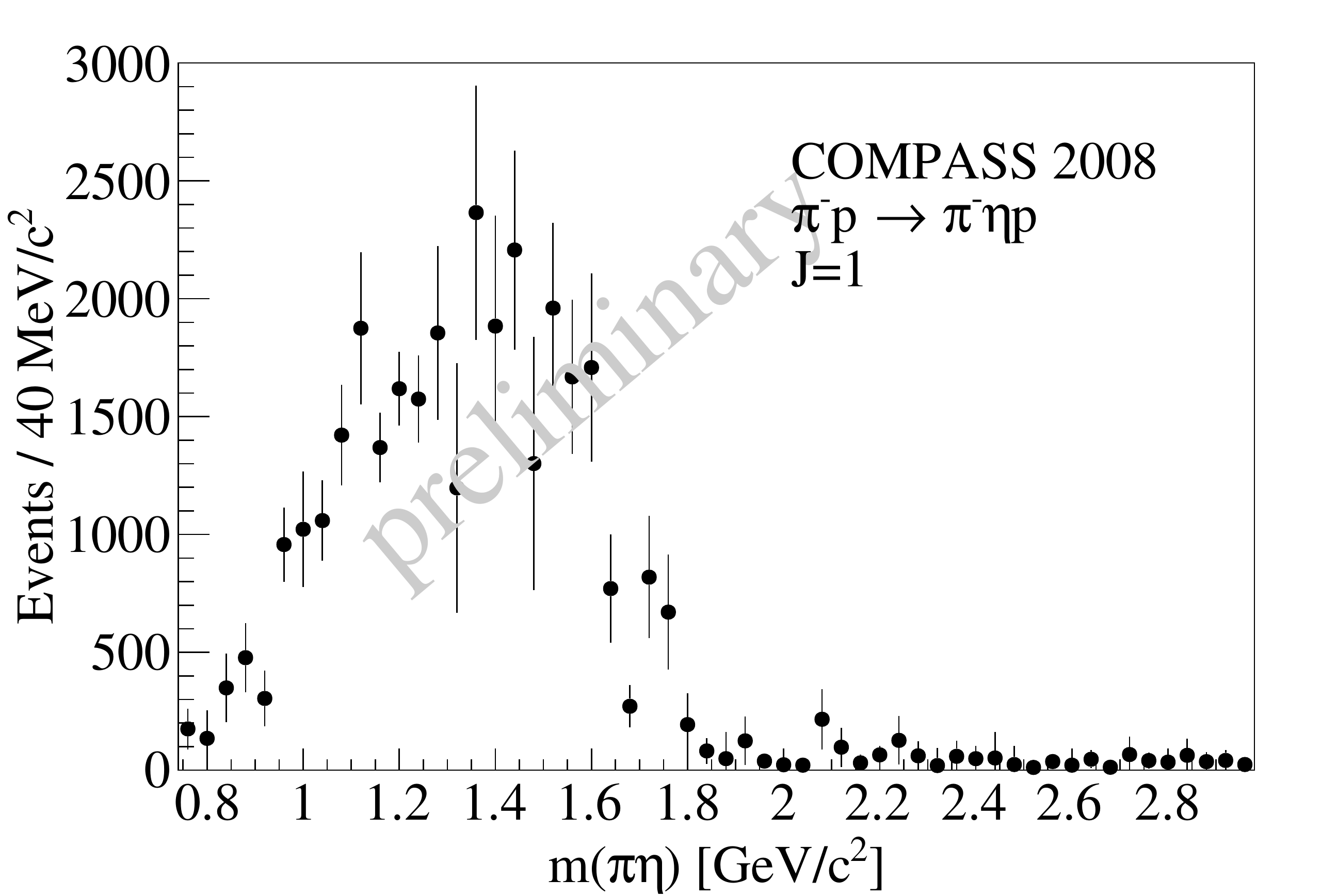}}
  \subfloat[$D$-wave, $J=2$]{\includegraphics[width=.32\textwidth]{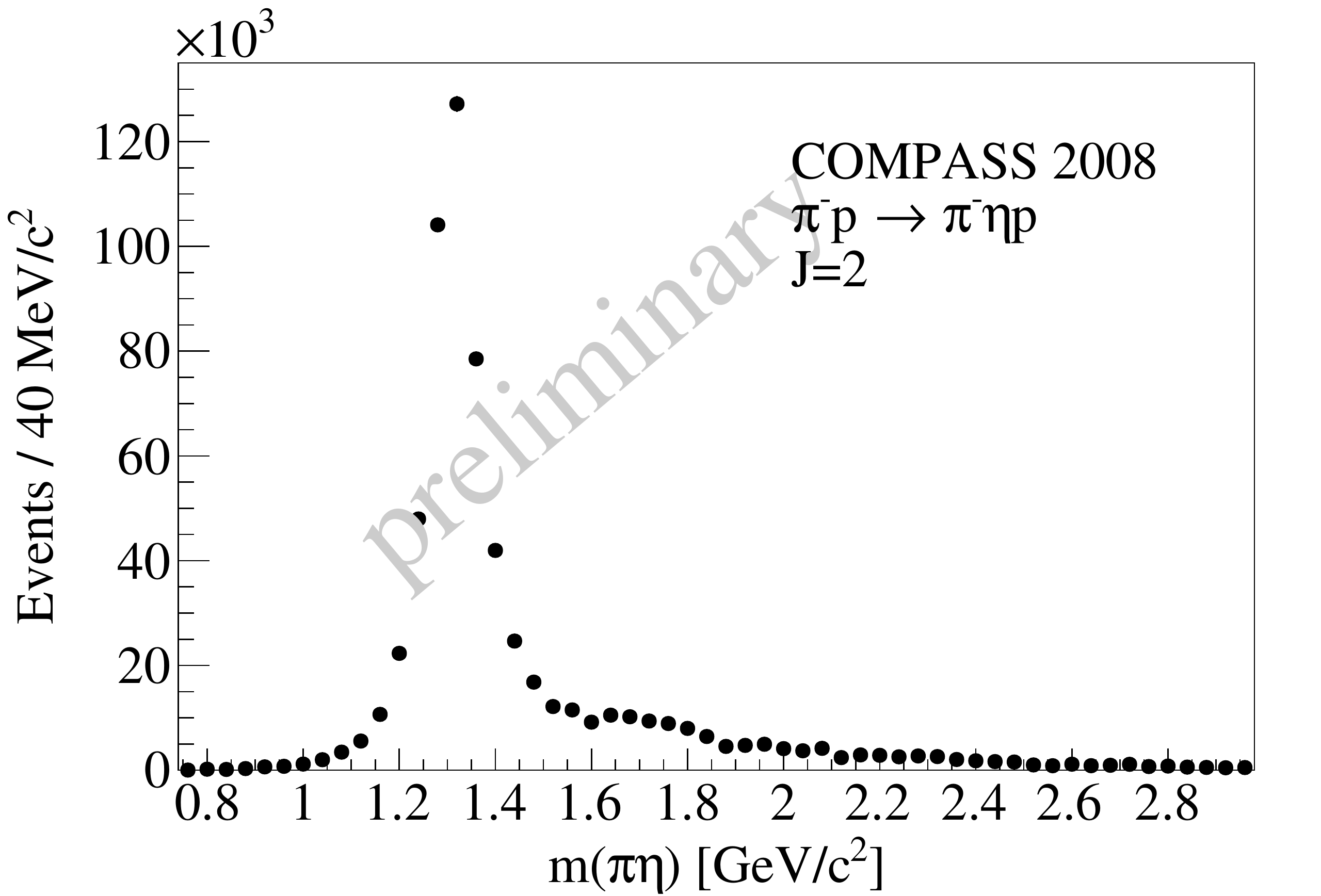}}
  \subfloat[$G$-wave,
  $J=4$]{\includegraphics[width=.32\textwidth]{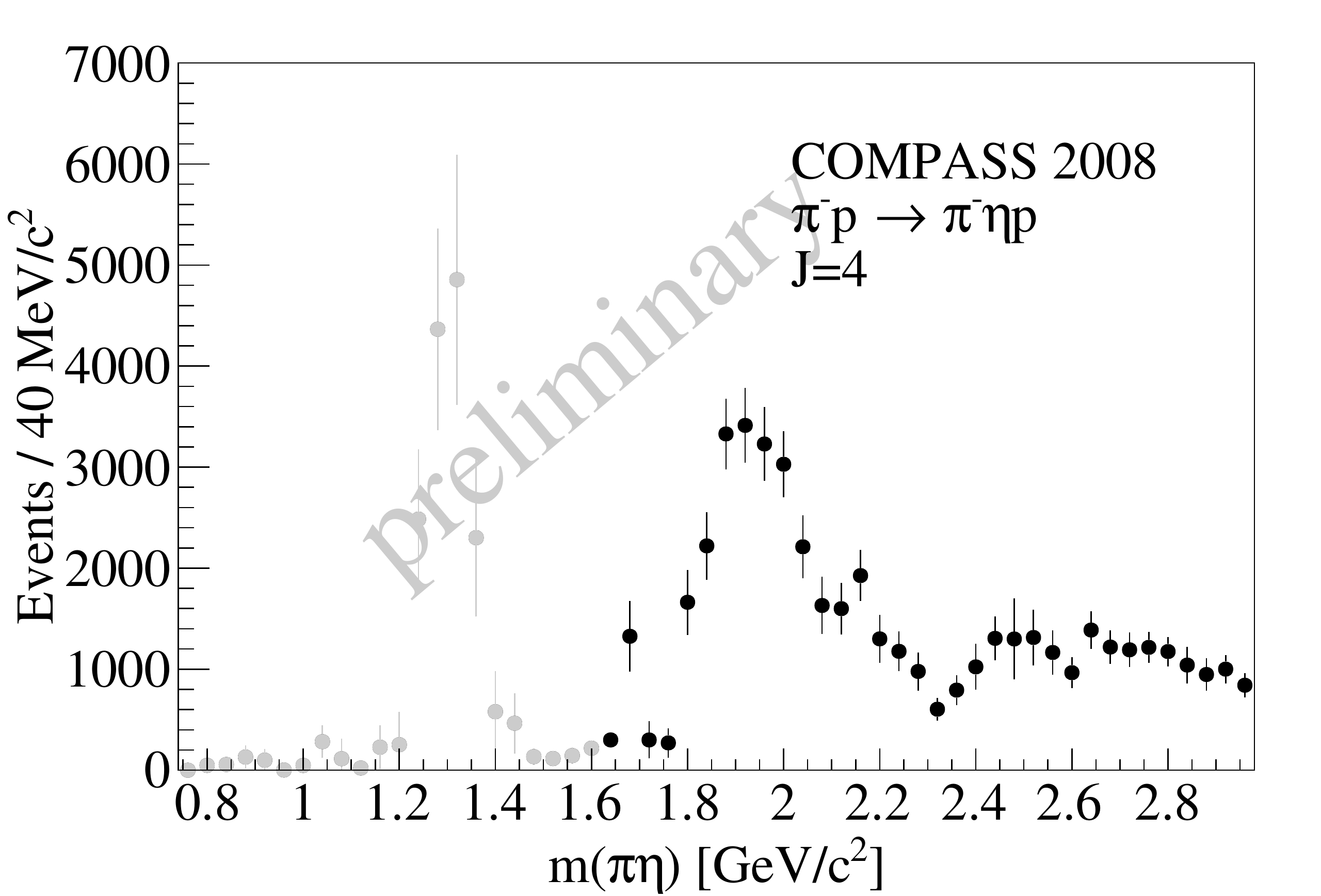}}
  \caption{Main waves of the $\eta(\to \pi^-\pi^+\gamma\gamma)\pi^-$ data.}
  \label{fig:pwa_etapi}
\end{figure}

We show the intensities of the main waves of the $\eta\pi^-$ data in
Fig.~\ref{fig:pwa_etapi}.  The $J=1$ $P$-wave shows a broad bump and
vanishing intensity above $1.8\,\textrm{GeV}$.  The $J=2$ wave is
dominated by the well-known $a_2(1320)$ resonance with a shoulder at
high mass.  Besides leakage from the dominant $J=2$ wave, the $J=4$
wave exhibits a clear $a_4(2040)$ signal, followed by a broad
structure at high mass.

\begin{figure}
  \centering
  \subfloat[$P$-wave, $J=1$]{\includegraphics[width=.32\textwidth]{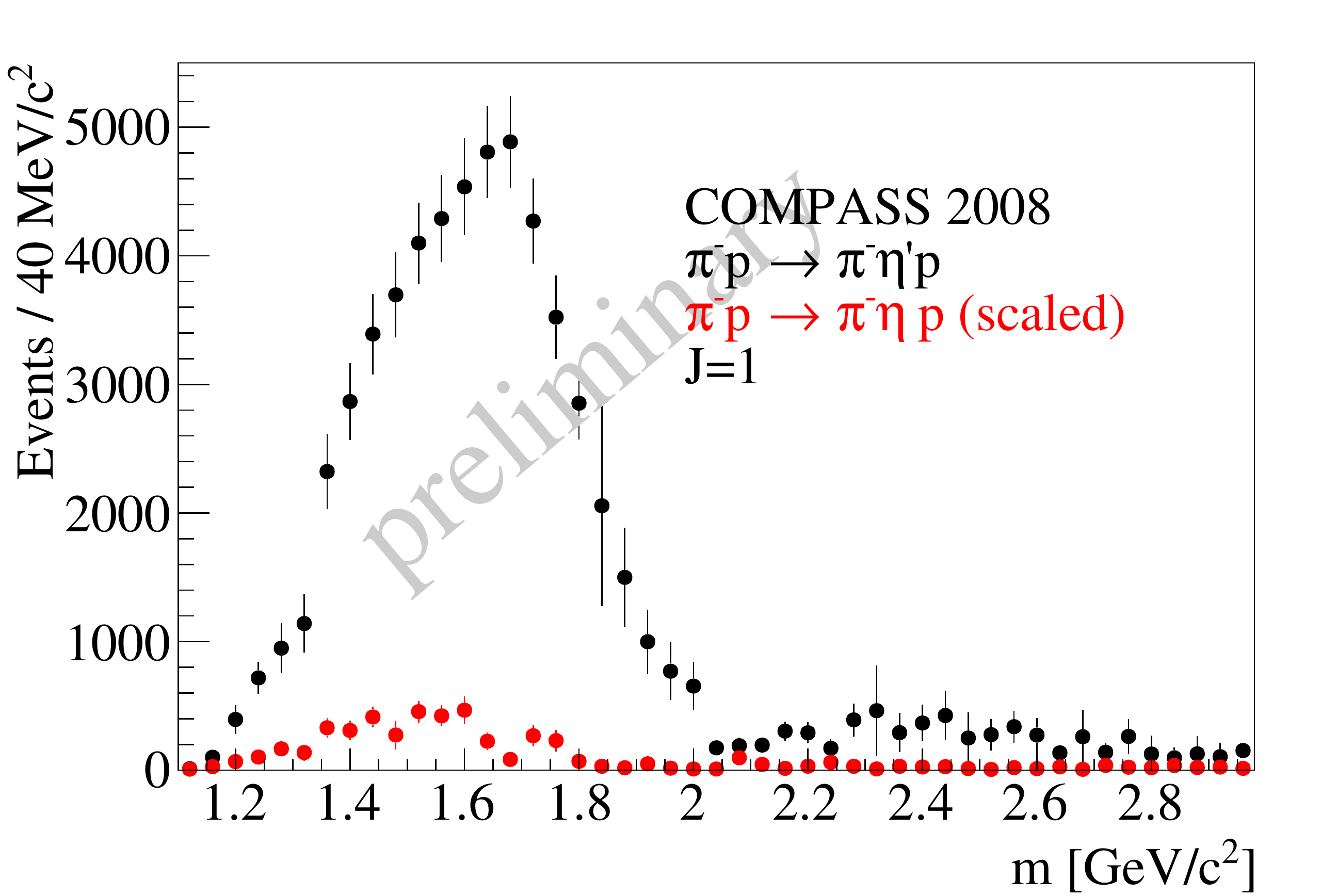}}
  \subfloat[$D$-wave, $J=2$]{\includegraphics[width=.32\textwidth]{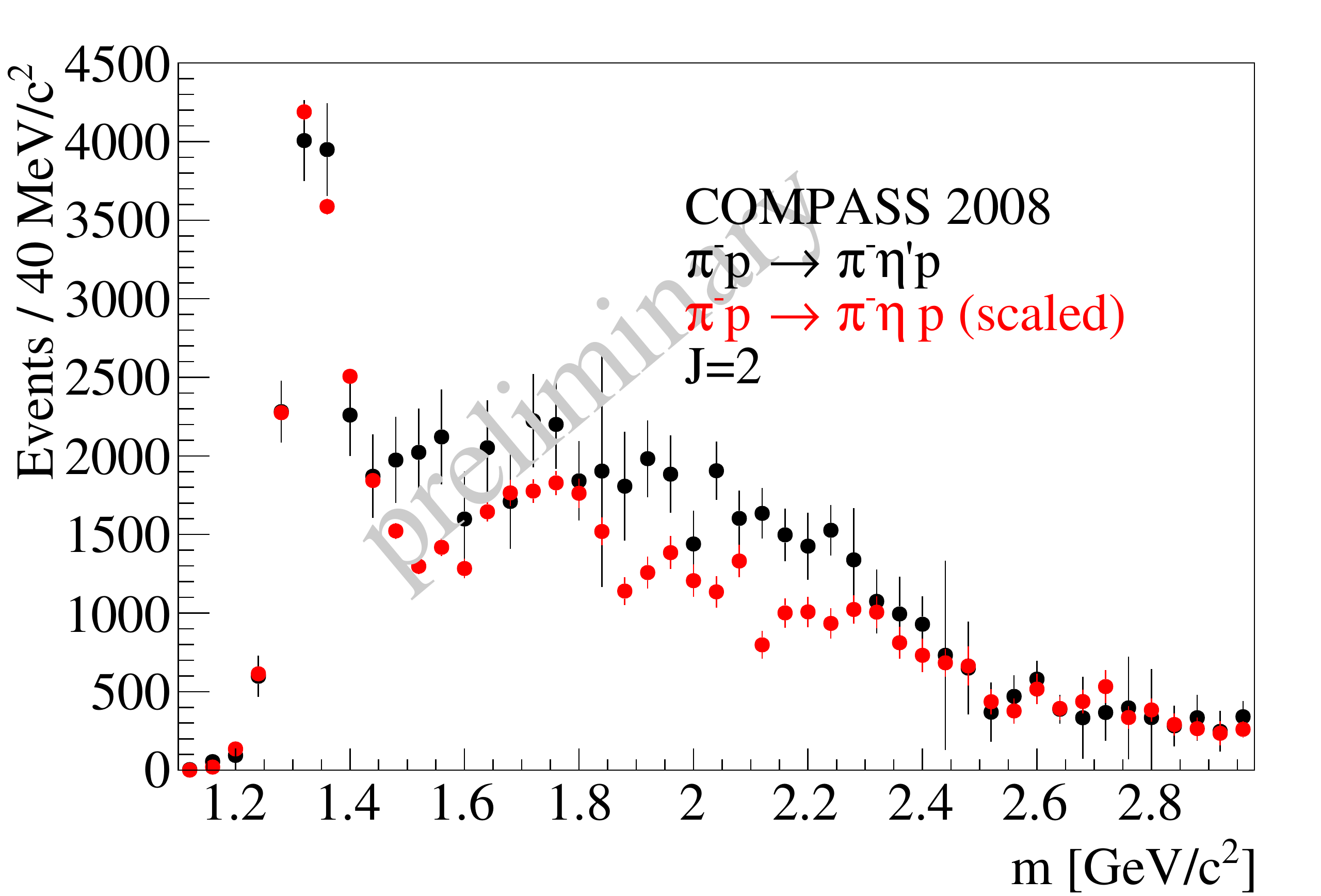}}
  \subfloat[$G$-wave, $J=4$]{\includegraphics[width=.32\textwidth]{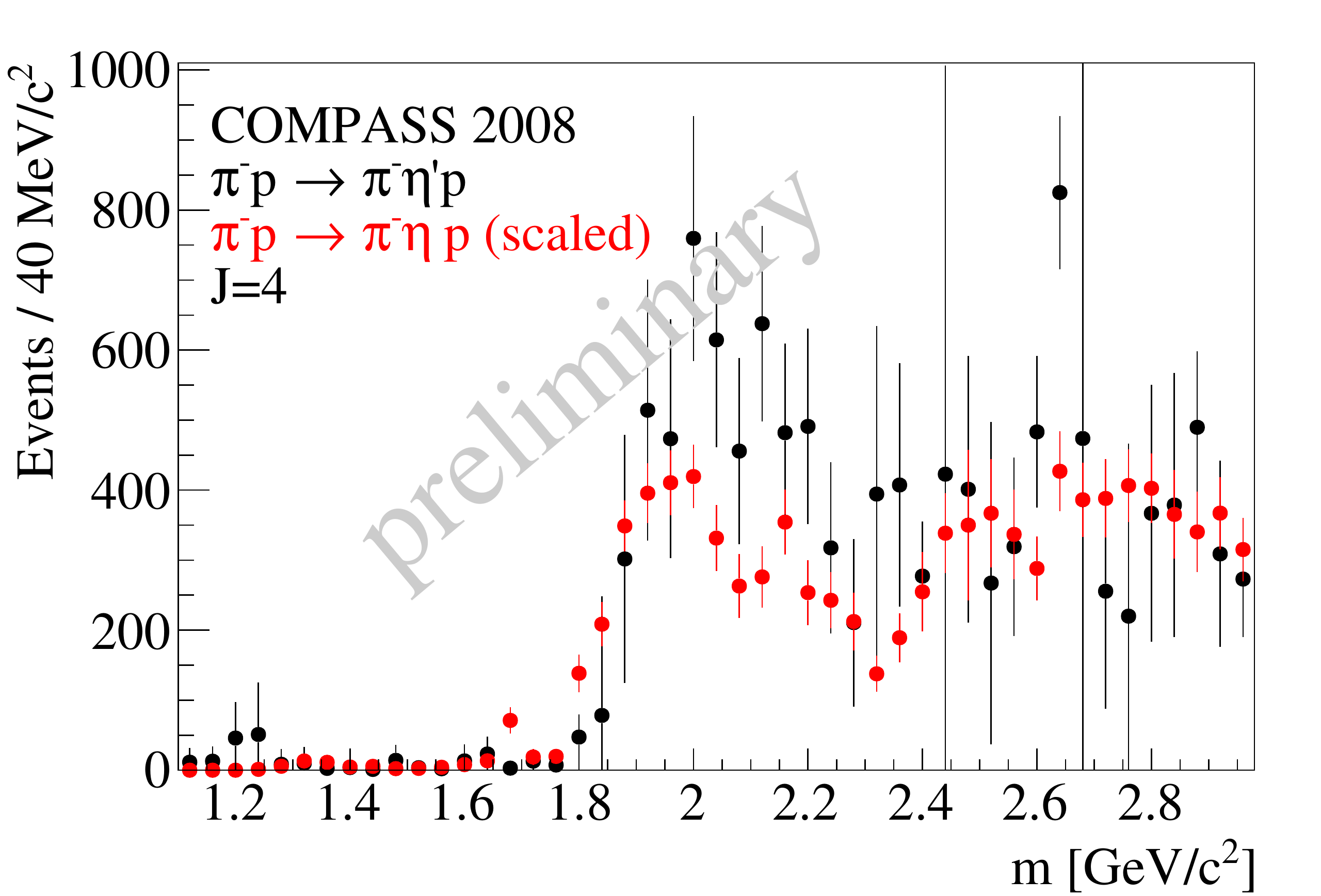}}
  \caption{Main waves of the $\eta'(\to \pi^-\pi^+\gamma\gamma)\pi^-$
    data.  In \textcolor{red}{red}: the $\eta\pi^-$ data
    multiplied by the mass-dependent phase-space factor from
    Eq.~1.1, taking into account final-state branching
    fractions.}
  \label{fig:overlay_intensities}
\end{figure}

The same partial waves are depicted for the $\eta'\pi^-$ data in
Fig.~\ref{fig:overlay_intensities}.  Again, we see a broad structure
in the $J=1$ $P$-wave, this time vanishing near $2\,\textrm{GeV}$ with
some intensity reappearing at higher masses.  In the $J=2$ wave, the
relative height of the high-mass shoulder compared to the peak is
enhanced compared to the $\eta\pi⁻$.  Similarly, the peak in the $J=4$
wave stands out less in the $\eta'\pi^-$ data.  Overlaid on the
$\eta'\pi^-$ data are the $\eta\pi^-$ data from
Fig.~\ref{fig:pwa_etapi}, where the content of each bin has been
multiplied with the factor from Eq.~\ref{eq:scale} and a factor taking
into account the final-state decays
$\eta^{(\prime)}\to\pi^-\pi^+\gamma\gamma$~\cite{Beringer:2012zz}.  We
find a surprising difference between different partial waves: whereas
the even waves with $J=2, 4$ show very similar behavior, the odd
$J=1$ wave is relatively enhanced in the $\eta'\pi^-$ data.  These
properties extend also to the waves with spins $J=3$, $5$, $6$ (not
shown): odd waves are relatively enhanced in $\eta'\pi^-$, even waves
largely agree after phase-space multiplication.


\begin{figure}
  \centering
  \subfloat[Phase $(J=1)-(J=2)$]{\includegraphics[width=.32\textwidth]{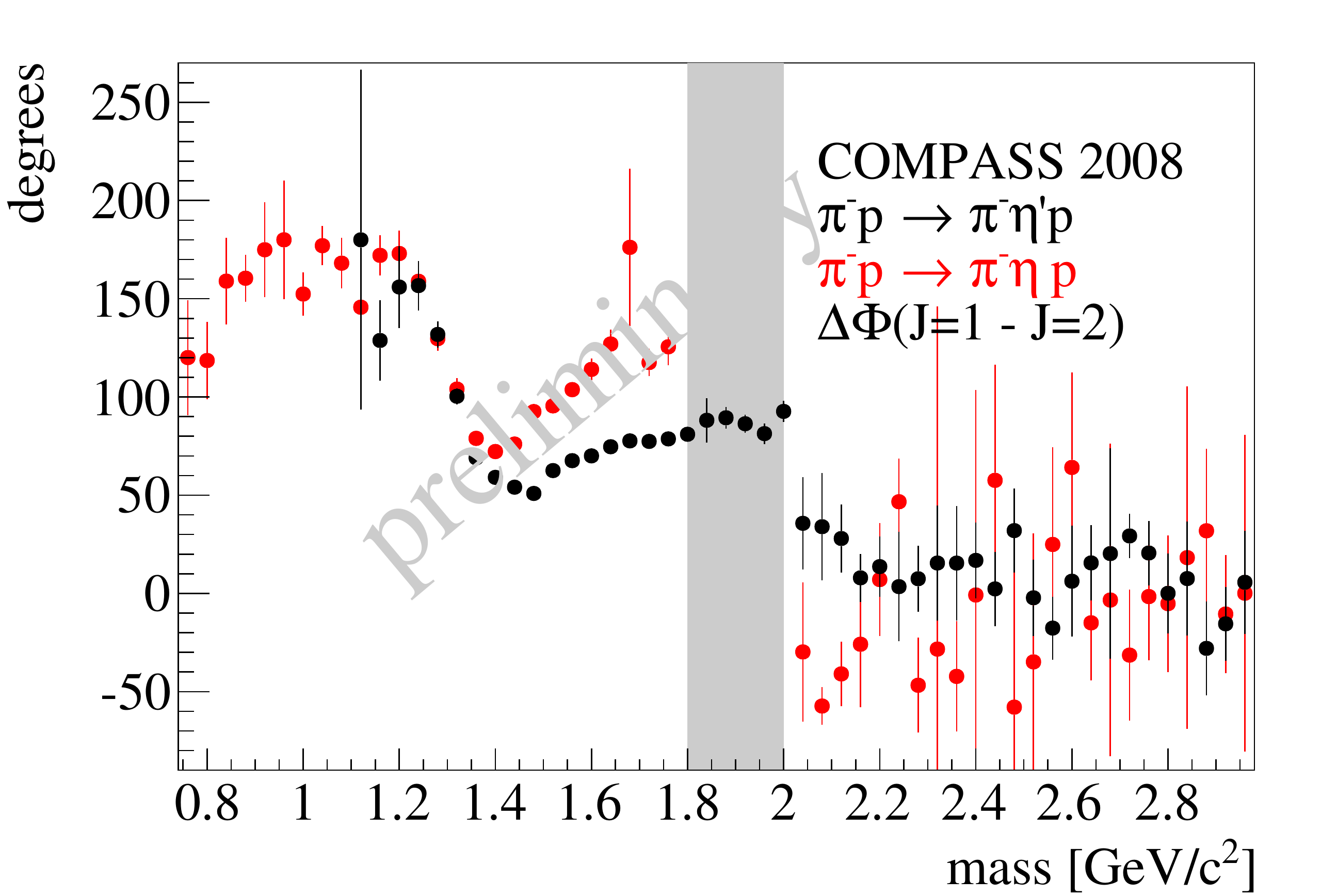}}
  \subfloat[Phase $(J=4)-(J=2)$]{\includegraphics[width=.32\textwidth]{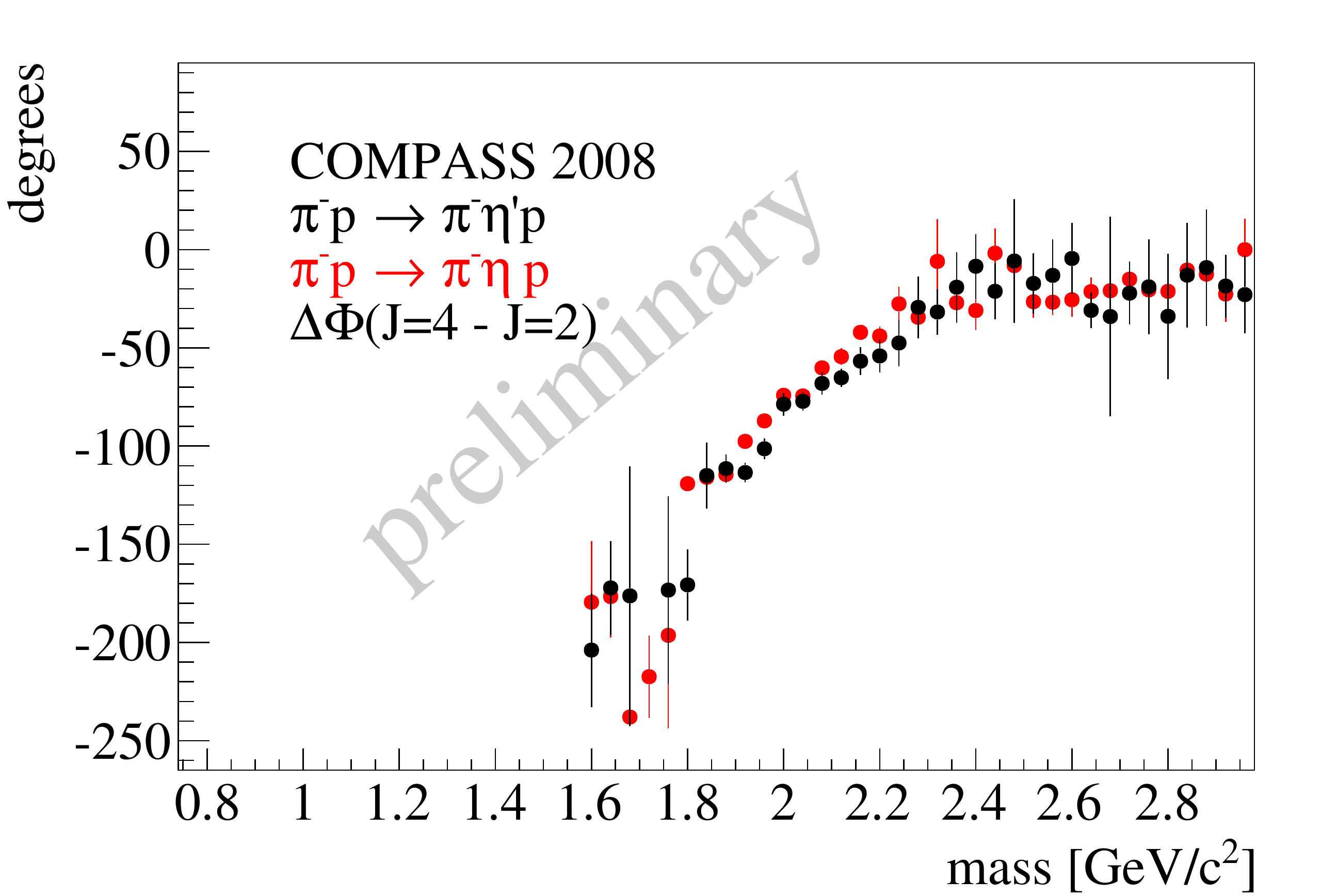}}
  \subfloat[Phase $(J=3)-(J=2)$]{\includegraphics[width=.32\textwidth]{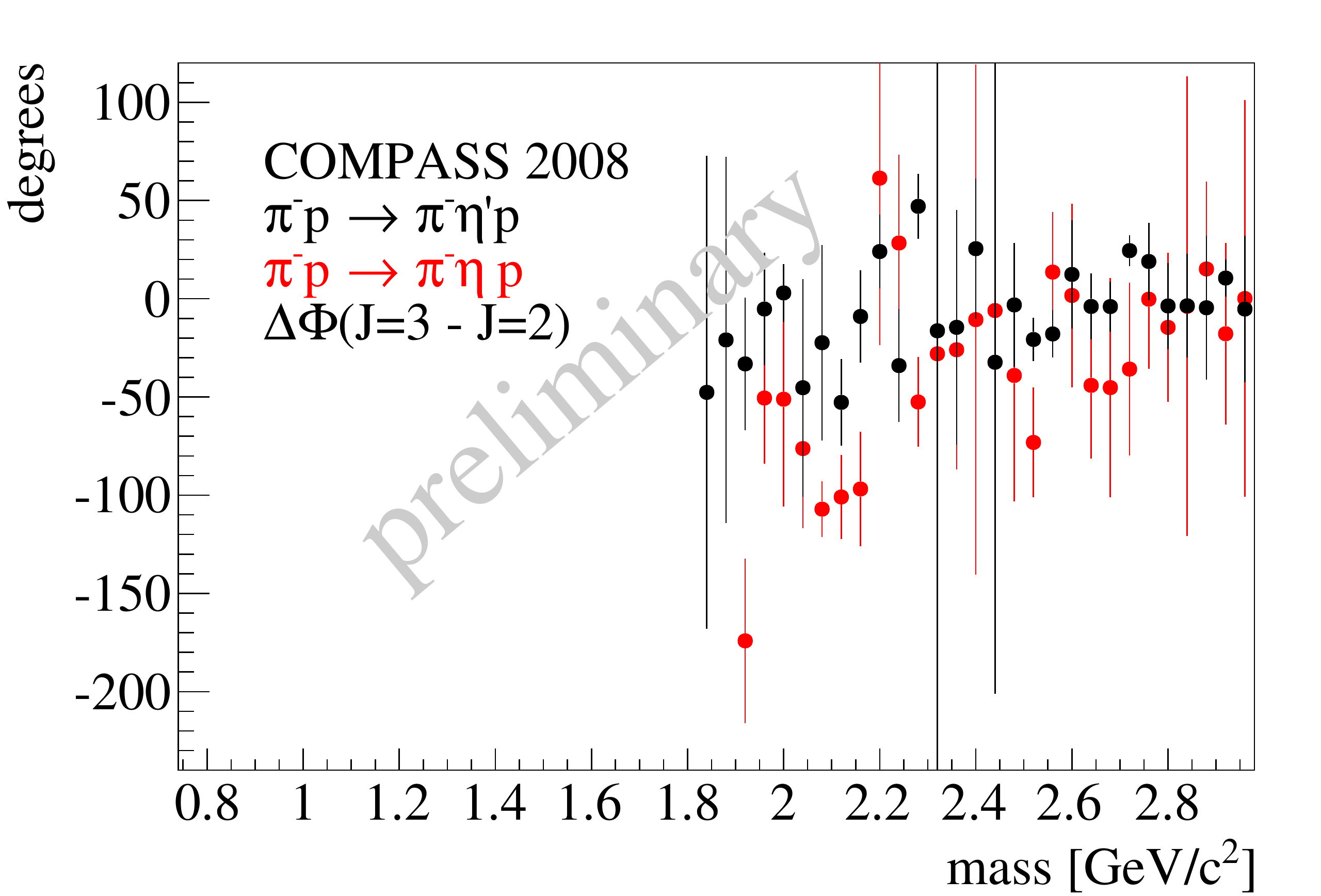}}
  \caption{Relative phases in $\eta'\pi^-$ and $\eta\pi^-$
    (\textcolor{red}{red}) for selected partial waves.}
  \label{fig:phases}
\end{figure}

For the phases similar behavior is observed, shown in
Fig.~\ref{fig:phases}: the phases between the even-spin waves $J=2$
and $J=4$ agree between the two channels.  The phases between the
$J=1$ and $J=2$ partial waves disagree in the region of the $J=1$
intensity peaks.  A particularly intriguing feature is the agreement
of this $(J=1) - (J=2)$ phase near the $\eta'\pi^-$ threshold.

The difference in even/odd behavior is detailed in
Tab.~\ref{tab:intensities}, where we show the relative intensities of
the various partial waves and the ratios of their integrals after
phase-space scaling.

\begin{table}[htbp]
  \centering

  \caption{Relative intensities of the $J=1$ to $6$ and
    $J=2$, $M=2$ partial waves resulting from the PWA fits integrated over    
    the mass range up to $3\,\textrm{GeV}$.  Experimental acceptance is
    taken into account.  The total $\eta'\pi^-$ to $\eta\pi^-$
    intensity ratio in this mass range amounts to $0.19\pm 0.02$.
    $R_{\textrm{corr}}$, given in the third row, is the ratio of the
    integral of the red histogram to the integral of the black histogram.}
  \begin{tabular}{c|ccccccc}
        J & 1 & 2 & 3 & 4 & 5 & 6 \\
        \hline
        $\frac{I_J(\eta \pi^-)}{I_{\textrm{total}}(\eta\pi)}$ [\%]
          & 4.4 & 81.9 & 0.3 & 6.9 & 0.1 & 0.7 \\
        $\frac{I_J(\eta' \pi^-)}{I_{\textrm{total}}(\eta'\pi)}$ [\%]
          & 41.7 & 42.3 & 3.7 & 8.4 & 0.9 & 1.2  \\
        $R_{\textrm{corr}}$ & $0.17\pm 0.01$ & $0.94\pm 0.02$ & $0.16\pm 0.05$ & $0.83\pm 0.07$ & $0.15\pm 0.12$ & $0.68\pm 0.15$ 
  \end{tabular}

  \label{tab:intensities}
\end{table}


\begin{thebibliography}{7}
\providecommand{\natexlab}[1]{#1}
\providecommand{\url}[1]{\texttt{#1}}
\expandafter\ifx\csname urlstyle\endcsname\relax
  \providecommand{\doi}[1]{doi: #1}\else
  \providecommand{\doi}{doi: \begingroup \urlstyle{rm}\Url}\fi

\bibitem[Klempt and Zaitsev(2007)]{Klempt:2007cp}
E.~Klempt, A.~Zaitsev, \emph{Glueballs, Hybrids, Multiquarks. Experimental facts versus QCD
  inspired concepts}, \href{http://dx.doi.org/10.1016/j.physrep.2007.07.006}{\emph{Phys. Rept.}, 454:\penalty0 1--202, 2007}, {\tt \href{http://arxiv.org/abs/arXiv:0708.4016}{arXiv:0708.4016 [hep-ph]}}

\bibitem[Schl{\"u}ter et~al.(2012)Schl{\"u}ter, Ryabchikov, D{\"u}nnweber, and
  Faessler]{Schluter:2012re}
T.~Schl{\"u}ter, {\it et al.}, {\emph Resonances of the systems $\pi^-\eta$ and $\pi^-\eta'$ in the
  reactions $\pi^-p \to \pi^-\eta p$ and $\pi^-p \to \pi^-\eta' p$ at COMPASS}, in proceedings of \emph{QNP 2012} \pos{PoS(QNP2012)074}, {\tt \href{http://arxiv.org/abs/arXiv:1207.1076}{arXiv:1207:1076 [hep-ex]}}.

\bibitem[Schl{\"u}ter(2012)]{Schluter:2012}
T.~Schl{\"u}ter, \emph{The $\pi^-\eta$ and $\pi^-\eta'$ Systems in Exclusive 190 GeV
  $\pi^-p$ Reactions at COMPASS (CERN)},
\href{http://wwwcompass.cern.ch/compass/publications/theses/2012_phd_schlueter.pdf}{PhD thesis, Ludwig-Maximilians-Universit{\"a}t, M{\"u}nchen, 2012}.

\bibitem[Abbon et~al.(2007)]{Abbon:2007pq}
P.~Abbon, {\it et~al.}, \emph{The COMPASS Experiment at CERN}, \href{http://dx.doi.org/10.1016/j.nima.2007.03.026}{\emph{Nucl. Instrum. Meth.}, A577:\penalty0 455--518, 2007}, {\tt [\href{http://arxiv.org/abs/hep-ex/0703049}{hep-ex/0703049}]}; M.G. Alekseev {\it et~al.}, {The COMPASS 2008 Spectrometer}, to be submitted, 2014.

\bibitem[Schl{\"u}ter et~al.(2011)Schl{\"u}ter, D{\"u}nnweber, Dhibar,
  Faessler, Geyer, Rajotte, Roushan, and W{\"o}hrmann]{Schluter:2011}
T.~Schl{\"u}ter, {\it et al.}, \emph{Large-Area Sandwich Veto Detector with WLS Fibre Readout for Hadron
  Spectroscopy at COMPASS}, \href{http://dx.doi.org/10.1016/j.nima.2011.05.069}{\emph{Nucl. Instrum. Meth.}, A654:\penalty0 219-224, 2011}, {\tt \href{http://arxiv.org/abs/arXiv:1108.4587}{arXiv:1108.4587 [physics.ins-det]}}.

\bibitem[Beringer et~al.(2012)]{Beringer:2012zz} J.~Beringer {\it
    et~al.}, \emph{Review of Particle Physics}, \newblock
  \href{http://dx.doi.org/10.1103/PhysRevD.86.010001}{\emph{Phys.Rev.},
      D86:\penalty0 010001, 2012}.

\end{thebibliography}

\end{document}